\documentclass{article}
\usepackage{spconf,amsmath,graphicx,settings, comment, dblfloatfix, subcaption, hyperref}
\hypersetup{hidelinks}
\immediate\write18{mkdir Graph_Tikz}  
\title{Real-time modelling of observation filter in the remote microphone technique for an active noise control application}
\name{
    Chung Kwan Lai,
    Bhan Lam, 
    Dongyuan Shi, 
    \href{mailto:EWSGAN@ntu.edu.sg}{Woon-Seng Gan}
\thanks{This research is supported by the Singapore Ministry of National Development and the National Research Foundation, Prime Minister's Office under the Cities of Tomorrow Research Programme (Award No.\@ COT-V4-2019-1). 
Any opinions, findings and conclusions or recommendations expressed in this material are those of the authors and do not reflect the view of National Research Foundation, Singapore, and Ministry of National Development, Singapore. \newline
We thank Kenneth Ooi for assistance with the mathematical derivation.}}

\address{School of Electrical and Electronic Engineering, Nanyang Technological University,  Singapore.}
\begin{document}
%
\maketitle

\begin{abstract}
The remote microphone technique (RMT) is often used in active noise control (ANC) applications to overcome design constraints in microphone placements by estimating the acoustic pressure at inconvenient locations using a pre-calibrated observation filter (OF), albeit limited to stationary primary acoustic fields. While the OF estimation in varying primary fields can be significantly improved through the recently proposed source decomposition technique, it requires knowledge of the relative source strengths between incoherent primary noise sources. This paper proposes a method for combining the RMT with a new source-localization technique to estimate the source ratio parameter. Unlike traditional source-localization techniques, the proposed method is capable of being implemented in a real-time RMT application. Simulations with measured responses from an open-aperture ANC application showed a good estimation of the source ratio parameter, which allows the observation filter to be modelled in real-time.
\end{abstract}

\section{Introduction}
\label{sec:intro}
Virtual sensing (VS) algorithms utilise physical remote monitoring sensors to estimate the acoustic pressure at a virtual position, and thus it is often employed in active noise control (ANC) applications with design constraints in error microphone placements where control is desired, such as at the human ears. \citep{Moreau2008, Edamoto2016, Deng2018, Sun2022}. One of the prominent VS methods is the remote microphone technique (RMT) \citep{Elliott2015}, which directly estimates the acoustic pressure at these virtual positions through pre-calibrated observation filters (OF), as shown in \fref{fig:block_diagram_RMT_noDelay}. This technique as described in \sref{sec:formulation_rmt}, however, assumes a stationary acoustic field to achieve a robust estimation at the virtual locations with the pre-calibrated fixed-coefficient OFs \citep{Elliott2020, Zhang2021}. This limits its application to where the acoustic field remains relatively stationary throughout the active control period, such as in road noise ANC in automobile cabins, where head-tracking techniques were applied with the RMT to continuously update the location of the virtual error microphones due to head movement \citep{Jung2018, Elliott2018a}. In cases where noise sources are time-varying and could arise from unknown directions, such as in the active control of noise through an open-aperture \citep{Shi2020, Lam2020} or mobile phones \citep{Cheer2018}, estimation performance will be degraded. While it was shown previously that the RMT estimation performance can be improved by reconstructing the correlation matrices (CMs) between microphones based on the superposition of CMs associated with its respective incoherent noise source \citep{Lai2022}, the reconstruction requires knowledge of the relative source strengths between these incoherent noise sources. Whereas source-localization techniques, such as the deconvolution approach for the mapping of acoustic sources (DAMAS), inverse acoustic method, or CM fitting (CMF) \citep{Yardibi2010a, Nelson2000}, could be utilised to estimate the source strengths, none of these methods was suitable for a VS application due to its modelling assumptions as detailed in \sref{sec:formulation_sourcetrack}. Through an open-aperture ANC implementation \citep{Lai2022}, the significance of the source ratio parameter is first described in \sref{sec:effect_mismatch}, followed by the proposed algorithm in \sref{sec:formulation_sourcetrack} to obtain the source ratio parameter through a source localization method, and finally its verification by simulation in \sref{sec:SourceTrack_result}.

\section{The remote microphone technique} 
\label{sec:formulation_rmt}
\begin{figure}[t] \centering 
	\includegraphics[width=\linewidth]{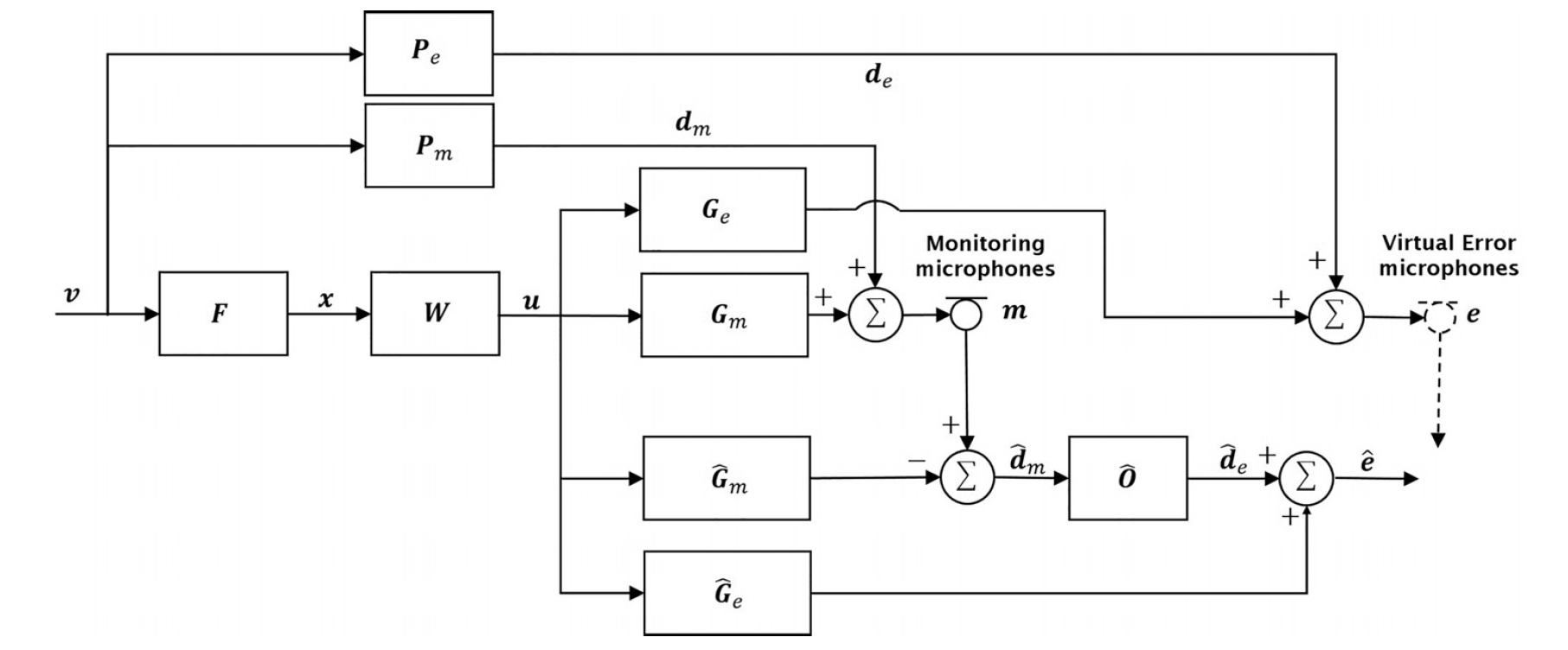}
	\caption{Block diagram of the virtual sensing control algorithm using remote microphone technique \citep{Jung2018}.}\label{fig:block_diagram_RMT_noDelay}
    \vspace{-0.5\baselineskip}
\end{figure}

\fref{fig:block_diagram_RMT_noDelay} shows the block diagram arrangement of a virtual sensing system using the remote microphone technique \citep{Moreau2008, Jung2018}. The observation filter $\boldsymbol{O}$ can be expressed in either the frequency domain or in the causally-constrained time domain to minimise the expected mean squared error between the estimated disturbance signal at the virtual microphone, $\hat{\boldsymbol{d}}_{e}$, and the actual disturbance signal, $\boldsymbol{d}_{e}$. Thus, the optimal observation filter in the frequency and causally-constrained time domain are expressed as \citep{Elliott2015, Jung2019}
{ \small
\begin{align} 
  \boldsymbol{O}_{opt}(j\omega) &= \boldsymbol{S}_{d_m d_e} \left(\boldsymbol{S}_{d_m d_m} + \beta\boldsymbol{I}\right)^{-1} \label{eqn:Oopt_freq_beta} \\
  &= \boldsymbol{P}_{e} \boldsymbol{S}_{vv} \boldsymbol{P}^{H}_{m} \left(\boldsymbol{P}_{m} \boldsymbol{S}_{vv} \boldsymbol{P}^{H}_{m} + \beta\boldsymbol{I} \right)^{-1} \nonumber
\end{align}
}%
and
{ \small
\begin{align} 
  \mathbf{O}_{opt} = \mathbf{R}_{d_m d_e}^{T}\left[0\right]\left(\mathbf{R}_{d_m d_m}\left[0\right]+\beta \mathbf{I}\right)^{-1}
 \label{eqn:Oopt_regularised_fir}
\end{align}
}%
respectively, where $\boldsymbol{P}_{e}$ and $\boldsymbol{P}_{m}$ are matrices of responses from the array of modelled primary sources $\boldsymbol{v}$ to the vectors of error and monitoring microphones $\boldsymbol{e}$ and $\boldsymbol{m}$, $E\left[ \cdot \right]$ is the expectation operator and $\beta$ is a regularisation parameter, $\boldsymbol{S}_{d_m d_e}$, $\boldsymbol{S}_{d_m d_m}$ and $\boldsymbol{S}_{vv}$ are the spectral density matrices and $\mathbf{R}_{d_m d_e}$, $\mathbf{R}_{d_m d_m}$ are the correlation matrices, each defined with a general notation of $\boldsymbol{S}_{xy}(\omega) = E\left[ \mathbf{y}(\omega) \mathbf{x}(\omega)^{H} \right]$ and $\mathbf{R}_{xy}[n_0] = E\left[ \mathbf{x}[n]\mathbf{y}[n-n_0]^{T}\right]$ respectively. For brevity, the $``\left[0\right]"$ notation from \eref{eqn:Oopt_regularised_fir} will be omitted throughout the paper, and while regularization can improve the robustness of the RMT when subject to uncertainty in the acoustic field \citep{Jung2018}, it is omitted to limit the scope (i.e. $\beta=0$). To evaluate the accuracy of the RMT, the overall estimation error of the RMT in the frequency domain is used, that is \citep{Jung2019}
{ \small
\begin{align}
  L_\epsilon = 10 \log_{10} \left| \frac{ \text{tr}\left\lbrace \boldsymbol{S}_{\epsilon \epsilon}\right\rbrace }{ \text{tr}\left\lbrace \boldsymbol{S}_{d_e d_e}\right\rbrace } \right| \label{eqn:EstimationError}
 \end{align}
 }%
 where $\epsilon = \boldsymbol{d}_{e} - \hat{\boldsymbol{d}}_{e} $ and $\text{tr}\{\cdot\}$ is the trace operator. 

\section{Source decomposition in the remote microphone technique}
\label{sec:effect_mismatch}
On the assumption of incoherence between modelled noise sources, it can be shown that the CM from \eref{eqn:Oopt_freq_beta} and \eref{eqn:Oopt_regularised_fir} can be further decomposed into a sum of CMs, with each associated to the respective noise sources \citep{Lai2022}. This allows the OF to be reconstructed in real-time based on the current primary acoustic field, given by
{ \footnotesize
\begin{align}
    \boldsymbol{O}_{opt}(j\omega) = \left( \sum_{n_v = 1}^{N_v} r_{n_v}^{2}\boldsymbol{S}_{d_m d_e}^{(n_v)} \right) \left[\left(\sum_{n_v = 1}^{N_v} r_{n_v}^{2}\boldsymbol{S}_{d_m d_m}^{(n_v)} \right) + \beta\boldsymbol{I}\right]^{-1} \label{eqn:Oopt_freq_decompose}
\end{align}
}%
and
{ \footnotesize
\begin{align}
    \boldsymbol{O}_{opt} = \left(\sum_{n_v = 1}^{N_v} r_{n_v}^{2}\mathbf{R}_{d_m d_e}^{(n_v)}\right)^{T}\left[\left(\sum_{n_v = 1}^{N_v} r_{n_v}^{2}\mathbf{R}_{d_m d_m}^{(n_v)}\right)+\beta \mathbf{I}\right]^{-1}, \label{eqn:Oopt_regularised_fir_decompose}
\end{align}
}%
where $N_v$ is the total number of the modelled primary sources in the system and $r_{n_v}$ denotes the source strength ratio at the $n_v$-th modelled primary source relative to its calibrated source strength.
\begin{figure}[!t] \centering
\includegraphics[width=0.9\linewidth]{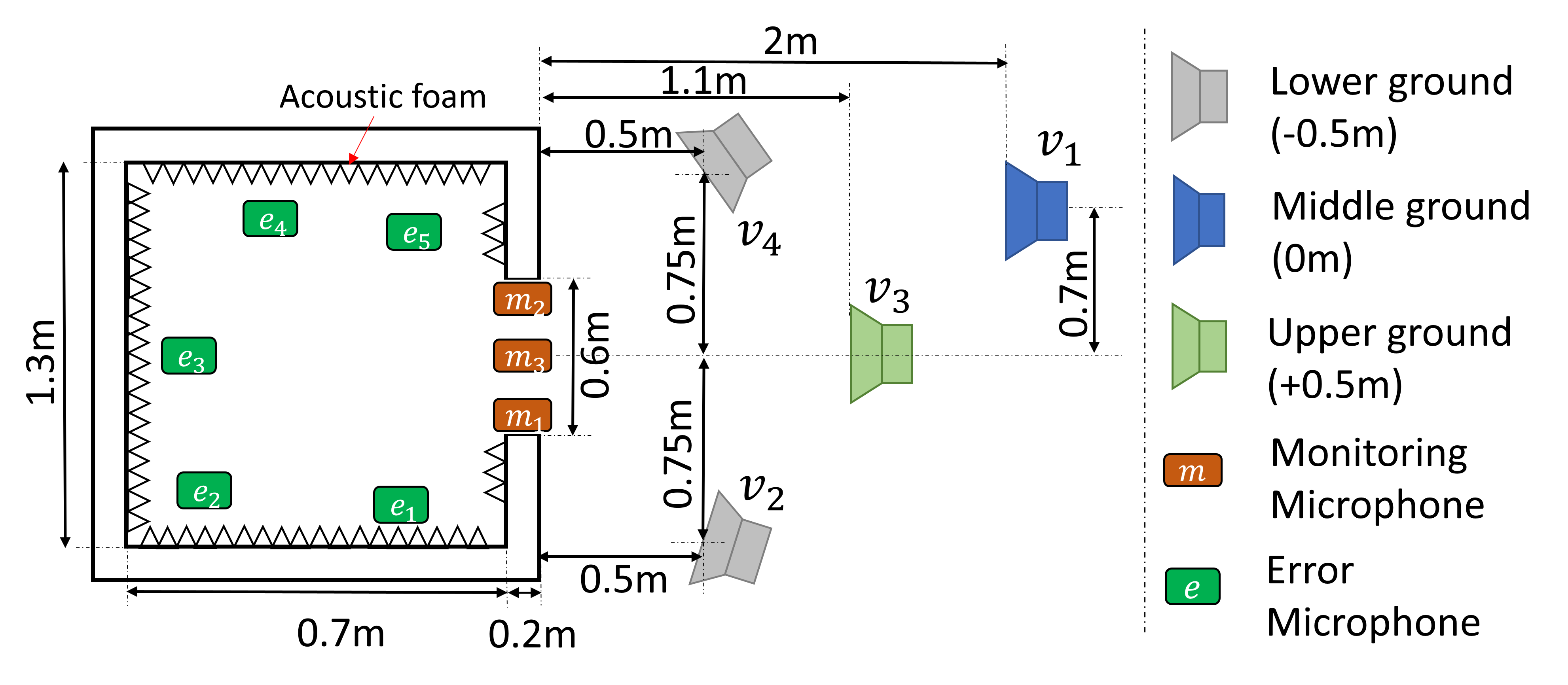}
\caption{Top view of the virtual sensing system for an open aperture, showing the arrangement of primary loudspeakers (Genelec 8302A) $v_1\;$--$\;v_4$, physical monitoring microphones $m_1\;$--$\;m_3$ and virtual error microphones $e_1\;$--$\;e_5$ (Pro Signal NPA415-OMNI). Signals were acquired and computed on a low-latency system (NI PXIe-8880) at a sampling frequency $F_s = 10 \text{kHz}$. \citep{Lai2022}.}\label{fig:Birdeye_placement}
    \vspace{-0.5\baselineskip}
\end{figure}
\newline \newline
As $r_{n_v}$ varies with time in a real-time implementation, it is important to understand the significance of this parameter. Additional real-time experiments from \citep{Lai2022} were conducted with its arrangement shown in \fref{fig:AmpCompare_Oopt}. Each loudspeaker reproduced white Gaussian noise during the calibration stage to obtain the individual CMs $\mathbf{R}_{d_m d_e}^{(n_v)}$ and $\mathbf{R}_{d_m d_m}^{(n_v)}$, which will be used to reconstruct the OF from \eref{eqn:Oopt_regularised_fir_decompose}. \fref{fig:AmpCompare_Group21_22_compare}--\ref{fig:AmpCompare_Group23_24_compare} showed the estimation error spectra when both $v_1$ and $v_2$ reproduced known amplitude ratios. While the nominal OF $\boldsymbol{O}_{opt}$ was directly obtained from both loudspeakers with the new amplitude ratio, the correctly estimated and mismatched OF $\hat{\boldsymbol{O}}$ and $\hat{\boldsymbol{O}}_{mis}$ uses the CM obtained from the calibration stage where the original amplitude was used for each individual loudspeaker, followed by the superposition formulation from \eref{eqn:Oopt_regularised_fir_decompose} using the correct and mismatched amplitude ratio input. The correctly estimated OF for both \fref{fig:AmpCompare_Group21_22_compare} and \ref{fig:AmpCompare_Group23_24_compare} showed a similar estimation spectra with the nominal OF which effectively validates \eref{eqn:Oopt_regularised_fir_decompose}. However, the estimation error can degrade when the wrong amplitude ratio is used. While \fref{fig:AmpCompare_Group21_22_compare} showed a decrease in estimation error at frequencies 400--600Hz, higher frequency region from 800--1000Hz did not show much change. \fref{fig:AmpCompare_Group23_24_compare} showed a larger difference in estimation error in a wider frequency range, suggesting a larger degradation in estimation error when its mismatch becomes larger. Thus, it can be concluded that the source ratio $r_{n_v}$ would play an important role to achieve robust estimation.
\begin{figure}[!t] \centering
\begin{subfigure}{0.49\linewidth}    \centering
\includegraphics[width=\linewidth]{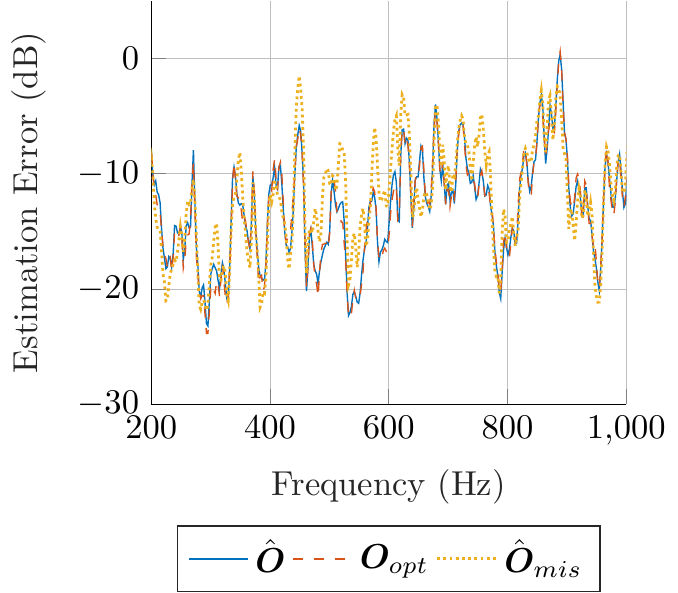}
\caption{$r_1 = 1.2, r_2 = 0.8$\\$r_{1, mis} = 0.8, r_{2, mis} = 1.2$}\label{fig:AmpCompare_Group21_22_compare} 
\end{subfigure}
\begin{subfigure}{0.49\linewidth} \centering
\includegraphics[width=\linewidth]{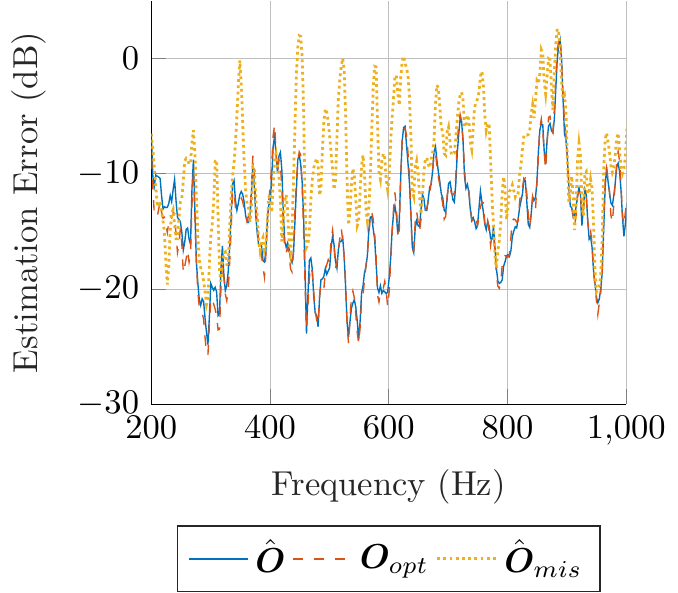}
\caption{$r_1 = 1.5, r_2 = 0.5$\\ $r_{1, mis} = 0.5, r_{2, mis} = 1.5$} \label{fig:AmpCompare_Group23_24_compare}
\end{subfigure}
\caption{The estimation spectra when loudspeaker 1 and 2 as arranged from \fref{fig:Birdeye_placement} \citep{Lai2022} were present, using the optimal OF $\boldsymbol{O}_{opt}$, correctly predicted OF  $\hat{\boldsymbol{O}}_{opt}$  and the mismatched OF $\hat{\boldsymbol{O}}_{mis}$ calculated from \eref{eqn:Oopt_regularised_fir_decompose}.} \label{fig:AmpCompare_Oopt}
\end{figure}

\section{Source tracking formulation}
\label{sec:formulation_sourcetrack}
In RMT implementation, multiple physical monitoring microphones are typically used, and thus the true cross-correlation matrix (CCM) between monitoring microphones $\mathbf{R}_{d_m d_m}$ can be obtained. By assuming incoherence between noise sources, and each CCM associated with that noise source was measured, the estimated CCM for $\hat{\mathbf{R}}_{d_m d_m}$ can be expressed as
{ \small
\begin{align}
    \hat{\mathbf{R}}_{d_m d_m} = \sum_{n_v = 1}^{N_v} z_{n_v} \mathbf{R}_{d_m d_m}^{(n_v)}, \label{eqn:Rmm_cross_decompose}
\end{align}
}%
with its respective source ratio parameter $z_{n_v} = r_{n_v}^2$ derived from \eref{eqn:Oopt_regularised_fir_decompose}. As $\mathbf{R}_{d_m d_m}$ changes with time due to changes in  $z_{n_v}$, the cost function can be formulated by minimizing its $l_2$ norm of the difference between the estimated CCM and its true CCM measured at that time. Thus, the minimization problem can be formulated as
{ \small
\begin{alignat}{2} 
    \min \; J(n)& = &&\left\| \mathbf{R}_{d_m d_m}(n) - \hat{\mathbf{R}}_{d_m d_m}(n)\right\|_2^2 \label{eqn:Cost_function_sourceTrack}\\
    s.t.& &&\quad \mathbf{a} \leq \mathbf{z} \leq \mathbf{b} \nonumber
\end{alignat}
}%
where $\mathbf{z} = [z_1 \; z_2 \; \cdots \; z_{N_v}]^T$ is the squared source ratio vector; $\mathbf{a}=[a_1 \; a_2 \cdots \; a_{N_v}]^T$ and $\mathbf{b}=[b_1 \; b_2 \cdots \; b_{N_v}]^T$ are the element-wise positive lower and upper bound vector constraints, i.e. $a_n, \; b_n \geq 0$ for all $n$. The objective function can thus be formulated with the use of the quadratic penalty function, defined as
{ \small
\begin{align}
    Q &= \left\|\mathbf{R}_{d_m d_m}-\mathbf{\hat{R}}_{d_m d_m}\right\|^2_2 \nonumber \\
    &+ \sum_{n_v=1}^{N_v}\sigma_{n_v} \left[h_{n_v}(z_{n_v}-a_{n_v})^{2} + g_{n_v}(z_{n_v}-b_{n_v})^{2}\right] \label{eqn:Quadratic_penalty_function}
\end{align}
}%
where $\mathbf{h} = [h_1 \; h_2 \; \cdots \; h_{N_v}]^T$ and $\mathbf{g} = [g_1 \; g_2 \; \cdots \; g_{N_v}]^T$ are penalty vectors that serves as a Heaviside function if the constraints were violated, given by 
{ \small
\begin{align}
    h_{n_v} = 
    \begin{cases}
        1, &  z_{n_v} < a_{n_v} \\
        0, & \text{otherwise}
    \end{cases}\; , \;
    g_{n_v}= 
    \begin{cases}
        1, &  z_{n_v} > b_{n_v} \\
        0, & \text{otherwise}
    \end{cases},
\end{align}
}%
and $\sigma_{n_v}$ is the penalty weight. Thus, the derivatives can be shown to be
{ \small
\begin{align}
    \frac{\partial Q}{\partial z_{n_v}} = &2 tr\left\lbrace -\mathbf{R}_{d_m d_m}\mathbf{R}_{d_m d_m}^{(n_v), T} 
    + \mathbf{R}_{d_m d_m}^{(n_v)}\mathbf{\hat{R}}_{d_m d_m}^{T} \right\rbrace\\
    + &2\sigma_{n_v}\left[(z_{n_v}-a_{n_v})h_{n_v} + (z_{n_v}-b_{n_v})g_{n_v}\right] \nonumber.
\end{align}
}%
Assuming that optimal $\mathbf{z}$ remained unconstrained, i.e. $\mathbf{a} < \mathbf{z}_{opt}< \mathbf{b} \text{, and thus } \mathbf{h} = \mathbf{g} = 0$, the optimal squared source ratio $\mathbf{z}_{opt}$ can be obtained by setting the derivatives to zero, that is
{ \small
\begin{align}
    \mathbf{z}_{opt} = \mathbf{A}^{-1}\mathbf{c},
    \label{eqn:zopt}
\end{align}
}%
where
{ \footnotesize
\begin{align}
\mathbf{A} = 
\begin{bmatrix}
\text{tr}\left\lbrace\mathbf{R}_{d_m d_m}^{(1)}\mathbf{R}_{d_m d_m}^{(1), T}\right\rbrace & \cdots & \text{tr}\left\lbrace\mathbf{R}_{d_m d_m}^{(1)}\mathbf{R}_{d_m d_m}^{(N_v), T}\right\rbrace \\
\text{tr}\left\lbrace\mathbf{R}_{d_m d_m}^{(2)}\mathbf{R}_{d_m d_m}^{(1), T}\right\rbrace &
\cdots & \text{tr}\left\lbrace\mathbf{R}_{d_m d_m}^{(2)}\mathbf{R}_{d_m d_m}^{(N_v), T}\right\rbrace \\
\cdots  & \ddots &\vdots \\ 
\text{tr}\left\lbrace\mathbf{R}_{d_m d_m}^{(N_v)}\mathbf{R}_{d_m d_m}^{(1), T}\right\rbrace &
\cdots & \text{tr}\left\lbrace\mathbf{R}_{d_m d_m}^{(N_v)}\mathbf{R}_{d_m d_m}^{(N_v), T}\right\rbrace
\end{bmatrix}
\end{align}
}%
and 
{ \footnotesize
\begin{align}
\mathbf{c} = 
\begin{Bmatrix}
\text{tr}\left\lbrace \mathbf{R}_{d_m d_m} \mathbf{R}_{d_m d_m}^{(1), T} \right\rbrace \\
\text{tr}\left\lbrace  \mathbf{R}_{d_m d_m} \mathbf{R}_{d_m d_m}^{(2), T} \right\rbrace \\
\vdots \\
\text{tr}\left\lbrace \mathbf{R}_{d_m d_m} \mathbf{R}_{d_m d_m}^{(N_v), T} \right\rbrace
\end{Bmatrix}.
\end{align}
}%
Since the optimal approach in \eref{eqn:zopt} is unconstrained and may lead to matrix ill-conditioning with large $N_v$ \citep{Yardibi2010}, an iterative gradient descent approach is adopted, whereby
{ \small
\begin{align}
    z_{n_v}^{(n+1)} = &z_{n_v}^{(n)} - \alpha_{n_v}\Biggl\lbrace \text{tr}\left\lbrace \mathbf{R}_{d_m d_m}^{(n_v)}\mathbf{\hat{R}}_{d_m d_m}^{T}-\mathbf{R}_{d_m d_m}\mathbf{R}_{d_m d_m}^{(n_v), T} \right\rbrace \nonumber \\
    + &\sigma_{n_v}\left[(z_{n_v}^{(n)}-a_{n_v})h_{n_v} + (z_{n_v}^{(n)}-b_{n_v})g_{n_v}\right] \Biggr\rbrace. \label{eqn:SourceRatio_gradientDescent}
\end{align}
}%
As $\mathbf{z}$ is being iterated closer to the optimal value, and therefore getting an accurate reconstruction of $\mathbf{R}_{d_m d_m}$, an accurate estimation of $\mathbf{R}_{d_m d_e}$ will be achieved indirectly as both correlation terms share the same source ratio term, which allows the reconstruction shown from \eref{eqn:Oopt_regularised_fir_decompose} to be implemented in real-time. The normalised estimation error of the source tracking technique across the $n$-th iteration can therefore be expressed given the general form of
{ \small
\begin{align}
    L_{xy}(n) = 10\log_{10}\left( \frac{\left\|\mathbf{R}_{xy}(n)-\hat{\mathbf{R}}_{xy}(n)\right\|^2_2}{\left\|\mathbf{R}_{xy}(n)\right\|^2_2}\right) \label{eqn:estimationError_sourceTrack}
\end{align}
}%
where the $xy$ subscript from \eref{eqn:estimationError_sourceTrack} can either be $d_m d_m$ or $d_m d_e$. 
\newline\newline\noindent
There is a distinction in this approach as compared to other traditional source-localization methods. Unlike the conventional CMF approach or DAMAS that makes use of a steering vector \citep{Yardibi2010a}, this method does not assume a free-field propagation from the noise source to the microphones which allows us to obtain the source ratio in a diffuse field. While it is certainly comparable to the acoustic inverse method \citep{Nelson2000}, this formulation strictly assumes that the noise source clusters are incoherent as explained in \sref{sec:effect_mismatch}. Additionally, the time domain CM is used instead of the cross-spectral density (CSD) in the frequency domain in this formulation which allows the causally constrained time-domain observation filter from \eref{eqn:Oopt_regularised_fir_decompose} to be reconstructed. This method, therefore, is well suited for certain virtual sensing applications, such as separating road noise and wind noise in automobile ANC as the CM caused by the road noise and wind noise can be measured separately.

\section{Simulation results}
\label{sec:SourceTrack_result}
\begin{figure}[!t] \centering
    \begin{subfigure}{0.49\linewidth}             \centering
        \includegraphics[width=\linewidth]{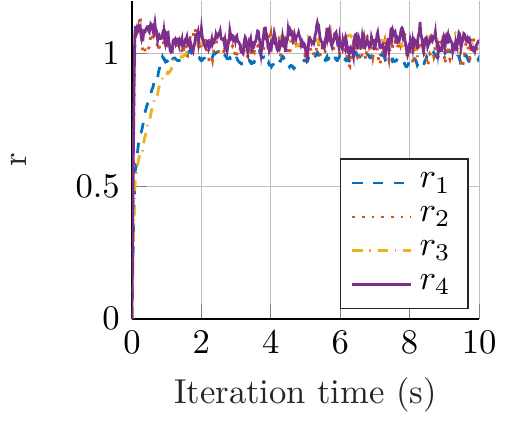}    \vspace{-1.5\baselineskip}
        \caption{}\label{fig:riterate_Group1} 
    \end{subfigure}
        \begin{subfigure}{0.49\linewidth} \centering
        \includegraphics[width=\linewidth]{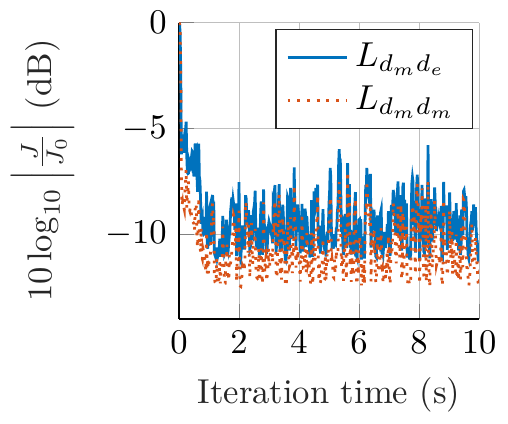}    \vspace{-1.5\baselineskip}
        \caption{} \label{fig:Jiterate_Group1}
    \end{subfigure}
    \caption{The plot of (\subref{fig:riterate_Group1}): Source ratio and (\subref{fig:Jiterate_Group1}): Source-tracking estimation error from \eref{eqn:estimationError_sourceTrack} over time using previous measurements from \citep{Lai2022}. $\alpha$ = 5, $\mathbf{a}=0$ and $\mathbf{b} = 5$ were used in this simulation, and the exact source-ratio used for each loudspeakers are $\mathbf{z}=1$. The true correlation matrix $\mathbf{R}_{d_m d_m}$ with the length of 400 is obtained for every time-frame of 200 samples (overlap of 50\%).} \label{fig:r_j_iterateSourceTrack} \vspace{-1\baselineskip}
\end{figure}
\begin{figure}[!t] \centering 
    \tikzsetnextfilename{Group1_estimationSourceTrack.tikz}
	\includegraphics[width=\linewidth, height=0.171\textheight]{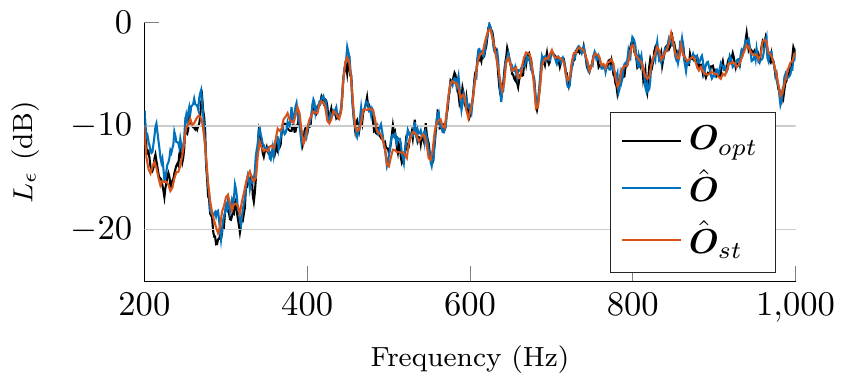}
	\caption{The estimation error spectra when all 4 loudspeakers from \fref{fig:Birdeye_placement} were present, using the optimal observation filter $\boldsymbol{O}_{opt}$, estimated observation filter $\hat{\boldsymbol{O}}$ with $\mathbf{z}=1$ and estimated observation filter with the use of source tracking iteration $\hat{\boldsymbol{O}}_{st}$.} \label{fig:EstimationError_SourceTrack}
\end{figure}

To verify the proposed algorithm in \sref{sec:formulation_sourcetrack}, simulations were made on the VS system in the case where all four primary loudspeakers from \fref{fig:Birdeye_placement} were present with a given source ratio parameter of 1, but the source ratio parameters were iterated along sample frame using \eref{eqn:SourceRatio_gradientDescent}. \fref{fig:riterate_Group1}--\subref{fig:Jiterate_Group1} show the iteration plot of the source ratio parameter and the normalised source-tracking estimation error over a period of 10 seconds using \eref{eqn:SourceRatio_gradientDescent} and \eref{eqn:estimationError_sourceTrack}. The iteration plot of the source ratio parameter showed convergence for all noise sources to around 1, with $r_1$ and $r_3$ converging quicker than $r_2$ and $r_4$. As the true $\mathbf{R}_{d_m d_m}$ with a length of 400 and overlap of 50\% has been obtained throughout the time frame, it is expected for the source ratio parameter to have random fluctuations from its stochastic nature, thus validating \eref{eqn:SourceRatio_gradientDescent} from obtaining its source ratio parameter. This convergence is also supported in \fref{fig:Jiterate_Group1} where $L_{d_m d_m}$ converges to around -13dB with expected random fluctuations. Interestingly, the estimation error of $\mathbf{R}_{d_m d_e}$ has decreased with time and converges at about -10dB even if it's indirectly estimated, which verifies the estimation concept described in \sref{sec:formulation_sourcetrack} where a good correlation in $L_{d_m d_m}$ and $L_{d_m d_e}$ is shown in \fref{fig:Jiterate_Group1}. Once $\mathbf{z}$ has been found through iteration, the observation filter due to source-tracking algorithm $\hat{\boldsymbol{O}}_{st}$ will be iteratively updated and was used to simulate the estimated error signals. It has been shown from the estimation error plot from \fref{fig:EstimationError_SourceTrack} that the estimation spectra between the nominal observation filter $\boldsymbol{O}_{opt}$ and source tracking observation filter $\hat{\boldsymbol{O}}_{st}$, which ultimately verifies the source-tracking algorithm.

\section{Conclusion}
\label{sec:Conclusion}
Although the remote microphone technique is sensitive to changes in the primary acoustic field, it can be reconstructed in real-time implementation through source decomposition as shown previously. The effect of the source ratio parameter on the estimation performance, however, is yet to be studied. In this paper, we have demonstrated the significance of the source ratio parameter when performing the RMT through source decomposition, as a large mismatch in the source ratio will cause a degradation of the estimation. Thus, we proposed a source-tracking algorithm by matching the CM directly and then iterating the source ratio parameter through gradient descent. To verify our algorithm, simulations with physical measurements from an open-aperture setup were conducted. Simulation results showed a good convergence when estimating the source ratio using the proposed algorithm, and thus can be used to reconstruct the observation filter in real-time implementation.

\bibliographystyle{IEEEbib}
\bibliography{0main}

\end{document}